\documentclass[a4paper,12pt]{article}
\usepackage{amsmath, amssymb, latexsym, amscd, amsthm,amsfonts,amstext}
\usepackage[mathscr]{eucal}
\usepackage{graphicx}
\usepackage{subfig}

\numberwithin{equation}{section}
\input{tcilatex}
\begin{document}

\title{Reconstruction formula for a 3-d phaseless inverse scattering problem
for the Schr\"{o}dinger equation}
\author{Michael V. Klibanov$^{\ast }$ and Vladimir G. Romanov$^{\circ }$
\and $^{\ast }$Department of Mathematics and Statistics \and University of
North Carolina at Charlotte \and Charlotte, NC 28223, USA \and $^{\circ }$%
Sobolev Institute of Mathematics, Novosibirsk 630090, Russia \and E-mails:
mklibanv@uncc.edu and romanov@math.nsc.ru}
\maketitle

\begin{abstract}
The inverse scattering problem of the reconstruction of the unknown
potential with compact support in the 3-d Schr\"{o}dinger equation is
considered. Only the modulus of the scattering complex valued wave field is
known, whereas the phase is unknown. It is shown that the unknown potential
can be reconstructed via the inverse Radon transform. Therefore, a long
standing problem posed in 1977 by K. Chadan and P.C.\ Sabatier in their book
\textquotedblleft Inverse Problems in Quantum Scattering Theory" is solved.
\end{abstract}


\textbf{AMS classification codes:} 65N15, 65N30, 35J25.

\graphicspath{{Figures/}}

%
%
%
%
%
%
%
%
%
%
%

%
%

\graphicspath{{FIGURES/}
{Figures/}
{FiguresJ/newfigures/}
{pics/}}

\section{Introduction}

\label{sec:1}

In this publication a long standing problem posed by Chadan and Sabatier in
1977 in chapter 10 of their classical book \cite{CS} is addressed. We
consider a 3-d inverse scattering problem for the Schr\"{o}dinger equation
with a compactly supported unknown potential in the frequency domain. Unlike
the common approach, we assume that only the modulus of the scattering field
is known, whereas the phase is unknown. The main result of this paper is a
reconstruction formula, which claims that this problem can be solved via the
inversion of the Radon transform. To the best knowledge of the authors, the
result of this paper represents the first rigorous reconstruction formula
for a phaseless inverse scattering problem without an assumption about
superpositions of signals caused by some separate targets, some of which are
known. It has been experienced by many people from their CT scans in
hospitals that the inverse Radon transform usually provides quite high
quality images.\ In this regard, we also refer to the book of Natterer \cite%
{Nat}. Therefore, the reconstruction formula of this paper paves the way for
future effective computations of some applied problems. An interesting
applied example is in imaging of nano structures, see section 4 in the paper
of Khachaturov \cite{Khach}.

The reason which has prompted Chadan and Sabatier to pose the phaseless
inverse scattering problem for the Schr\"{o}dinger equation in \cite{CS} is
that in the quantum scattering in the frequency domain one is measuring the
differential scattering cross section. The latter is the modulus of the
scattered complex valued wave field, see page 8 in the book of Newton \cite%
{Newton}. However, the phase is not measured. On the other hand, the entire
inverse scattering theory in the frequency domain is based on the assumption
that both the modulus and the phase are measured outside of the support of a
scatterer, see, e.g. books of Chadan and Sabatier \cite{CS}, Isakov \cite{Is}%
, Newton \cite{Newton} as well as papers of Novikov \cite{Nov1,Nov2}.

Because of a number of its important applications, there are many
publications about the problem of phase reconstruction. As some examples, we
refer to Aktosun and Sacks \cite{AS}, Berk and Majkrzak \cite{BM}, Dobson 
\cite{Dob}, Feinup \cite{F}, Gerth, Hofman, Birkholz, Koke and Steinmeyer 
\cite{Hofman}, Ivanyshyn, Kress and Serranho \cite{Iv1}, Ivanyshyn and Kress 
\cite{Iv2}, Ladd and Palmer \cite{Ladd}, Nazarchuk, Hryniv and Synyavsky 
\cite{NHS}, and Ruhlandt, Krenkel, Bartels and Salditt \cite{Ruhl}.

In the recent preprint of Novikov \cite{Nov3} another reconstruction formula
is obtained for the phaseless inverse scattering problem for the Schr\"{o}%
dinger equation. We now point to the main difference between our result and
the one of \cite{Nov3}. In the inversion formulae of Theorem 2.1 of \cite%
{Nov3} three measurements are considered: one from the unknown potential and
two more for the case when that target potential is complemented by two
other compactly supported potentials, which are known and whose supports do
not intercept with the support of the target potential. This means that
superpositions of signals scattered by three separate targets are considered
in \cite{Nov3}. On the other hand, we consider measurements of the modulus
of the scattered wave field generated only by a single compactly supported
potential. The inverse Radon transform is not used in \cite{Nov3}, and the
method of the proof of the main result (Theorem 1) here is significantly
different from the one in \cite{Nov3}.

In the recent work of Klibanov \cite{KSIAP} uniqueness theorems for the 3-d
phaseless inverse scattering problem for the Schr\"{o}dinger equation were
proved, also see the work \cite{AA} for a similar result for the acoustic
equation. Uniqueness of the reconstruction of a complex valued function with
compact support from the modulus of its Fourier transform was proved in \cite%
{K6,KK}. However, proofs in papers \cite{KSIAP}-\cite{KK} are not
constructive.

In section 2 we formulate the problem and the main result. In section 3 we
prove the main result. In section 4 we prove a certain lemma, which is
formulated in section 3.

\section{The Main Result}

\label{sec:2}

Let $B>0$ be a number and $\Omega =\left\{ \left\vert x\right\vert
<B\right\} \subset \mathbb{R}^{3}$ be the ball of the radius $B$ with the
center at $\left\{ 0\right\} $. Denote the corresponding sphere $S=\left\{
\left\vert x\right\vert =B\right\} .$ Let the potential $q\left( x\right)
,x\in \mathbb{R}^{3}$ be a real valued function such that 
\begin{equation}
q\left( x\right) \in C^{4}\left( \mathbb{R}^{3}\right) ,  \label{1.1}
\end{equation}%
\begin{equation}
q\left( x\right) \geq 0,\forall x\in \Omega ,  \label{1.2}
\end{equation}%
\begin{equation}
q\left( x\right) =0\text{ for }x\in \mathbb{R}^{3}\diagdown \Omega .
\label{1.3}
\end{equation}%
Let $x^{0}$ be the position of the point source. As the forward problem, we
consider the following 
\begin{equation}
\Delta _{x}u+k^{2}u-q\left( x\right) u=-\delta \left( x-x^{0}\right) ,\quad
x\in \mathbb{R}^{3},  \label{1.4}
\end{equation}%
\begin{equation}
u\left( x,x^{0},k\right) =O\left( \frac{1}{\left\vert x-x^{0}\right\vert }%
\right) ,\quad \left\vert x\right\vert \rightarrow \infty ,  \label{1.50}
\end{equation}%
\begin{equation}
\sum\limits_{j=1}^{3}\frac{x_{j}-x_{j}^{0}}{\left\vert x-x^{0}\right\vert }%
\partial _{x_{j}}u\left( x,x^{0},k\right) +iku\left( x,x^{0},k\right)
=o\left( \frac{1}{\left\vert x-x^{0}\right\vert }\right) ,\quad \left\vert
x\right\vert \rightarrow \infty .  \label{1.5}
\end{equation}%
Here the frequency $k>0$ and conditions (\ref{1.50}), (\ref{1.5}) are valid
for every fixed source position $x^{0}.$ Theorem 3.3 of the paper of
Vainberg \cite{V1}, Theorem 6 of Chapter 9 of the book of Vainberg \cite{V}
as well as Theorem 6.17 of the book of Gilbarg and Trudinger \cite{GT}
guarantee that for each pair $\left( k,x^{0}\right) \in \left( 0,\infty
\right) \mathbb{\times R}^{3}$ there exists a unique solution $u\left(
x,x^{0},k\right) $ of the problem (\ref{1.4}), (\ref{1.50}), (\ref{1.5})
such that it can be represented in the form%
\begin{equation}
u\left( x,x^{0},k\right) =u_{0}\left( x,x^{0},k\right) +u_{sc}\left(
x,x^{0},k\right) ,  \label{1.6}
\end{equation}%
\begin{equation}
u_{0}=\frac{\exp \left( -ik\left\vert x-x^{0}\right\vert \right) }{4\pi
\left\vert x-x^{0}\right\vert },\text{ }u_{sc}\in C^{4}\left( \left\{
\left\vert x-x^{0}\right\vert \geq \eta \right\} \right) ,\forall \eta
>0,\forall \beta \in \left( 0,1\right) .  \label{1.7}
\end{equation}

For any number $a\in \mathbb{R}$ consider the plane $P_{a}=\left\{
x_{3}=a\right\} .$ Consider the disk $Q_{a}=\overline{\Omega }\cap P_{a}$
and let $S_{a}=S\cap P_{a}$ be its boundary. Clearly $Q_{a}\neq \varnothing $
for $a\in \left( -B,B\right) $ and $Q_{a}=\varnothing $ for $\left\vert
a\right\vert \geq B.$ Denote $0_{a}=\left( 0,0,a\right) \in Q_{a}$ the
orthogonal projection of the origin on the plane $P_{a}.$ We have 
\begin{equation*}
\Omega =\dbigcup\limits_{a=-B}^{B}Q_{a},\partial \Omega
:=S=\dbigcup\limits_{a=-B}^{B}S_{a}.
\end{equation*}%
In our inverse problem we assume that the modulus $\left\vert
u_{sc}\right\vert $ of the scattered wave is measured for all pairs $x^{0},x$
running along the circle $S_{a}$ for every $a\in \left( -R,R\right) $ and
for all frequencies $k>0$.

\textbf{Phaseless Inverse Scattering Problem}. \emph{Suppose that the
potential }$q\left( x\right) $\emph{\ satisfies conditions (\ref{1.1})-(\ref%
{1.3}). Determine the function }$q\left( x\right) $\emph{\ for }$x\in \Omega
,$\emph{\ assuming that the following function }$f\left( x,x^{0},k\right) $%
\emph{\ is known }%
\begin{equation}
f\left( x,x^{0},k\right) =\left\vert u_{sc}\left( x,x^{0},k\right)
\right\vert ,\forall x^{0},x\in S_{a},x\neq x^{0},\forall a\in \left(
-B,B\right) ,\forall k\in \left( 0,\infty \right) .  \label{1.8}
\end{equation}

\textbf{Remark 1}. As to the issue of collecting experimental data, it
follows from (\ref{1.8}) and Theorem 1 that if one wants to image only one
2-d cross-section $Q_{a}$ of the potential $q,$ then it is sufficient to run
independently both sources $x^{0}$ and detectors $x$ only around the circle $%
S_{a}$. This is more economical than running them independently around the
entire sphere $S$.

For an arbitrary $a\in \left( -B,B\right) $ and for any pair of points $%
x^{0},x\in S_{a}$ let $L\left( x,x^{0}\right) $ be the interval of the
straight line connecting them. Denote $B_{a}=\sqrt{B^{2}-a^{2}}$ the radius
of the circle $S_{a}.$ Since our reconstruction formula is based on the
inversion of the two-dimensional\ Radon transform, we now parametrize $%
L\left( x,x^{0}\right) $ in the conventional parametrization of the Radon
transform \cite{Nat}. Let $n$ be the unit normal vector to the line $L\left(
x,x^{0}\right) $ lying in the plane $P_{a}$ and pointing outside of the
point $0_{a}.$ Let $\alpha \in \left( 0,2\pi \right] $ be the angle between $%
n$ and the $x_{1}-$axis. Then $n=n\left( \alpha \right) =\left( \cos \alpha
,\sin \alpha \right) $ (it is convenient here to discount the third
coordinate of $n,$ which is zero). Let $s$ be the signed distance of $%
L\left( x,x^{0}\right) $ from the point $0_{a}$ (\cite{Nat}, page 9). It is
clear that there is a one-to-one correspondence between pairs $\left(
x,x^{0}\right) $ and $\left( n\left( \alpha \right) ,s\right) ,$%
\begin{equation}
\left( x,x^{0}\right) \Leftrightarrow \left( n\left( \alpha \right)
,s\right) ;x,x^{0}\in S_{a}\in S_{a},\alpha =\alpha \left( x,x^{0}\right)
\in \left( 0,2\pi \right] ,s=s\text{ }\left( x,x^{0}\right) \in \left(
-B_{a},B_{a}\right) .  \label{1.80}
\end{equation}%
Hence, we can write 
\begin{equation}
L\left( x,x^{0}\right) =\left\{ y_{a}=\left( y_{1},y_{2},a\right)
:\left\langle y,n\left( \alpha \right) \right\rangle =s\right\} ,
\label{1.81}
\end{equation}%
where $y=\left( y_{1},y_{2}\right) \in \mathbb{R}^{2},\left\langle
,\right\rangle $ is the scalar product in $\mathbb{R}^{2}$ and parameters $%
\alpha =\alpha \left( x,x_{0}\right) $ and $s=s$ $\left( x,x_{0}\right) $
are defined as in (\ref{1.80}).

Consider an arbitrary function $g=g\left( y\right) \in C^{4}\left(
P_{a}\right) $ such that $g\left( y\right) =0$ for $y\in $ $P_{a}\diagdown
Q_{a}.$ Hence, 
\begin{equation}
\dint\limits_{L\left( x,x^{0}\right) }g\left( y\right) d\sigma
=\dint\limits_{\left\langle y,n\left( \alpha \right) \right\rangle
=s}g\left( y\right) d\sigma ,  \label{1.82}
\end{equation}%
for all $x,x^{0}\in S_{a},$ where $\alpha =\alpha \left( x,x^{0}\right) ,s=s$
$\left( x,x^{0}\right) $ as in (\ref{1.80}). In (\ref{1.82}) $\sigma $ is
the arc length and the parametrization of $L\left( x,x^{0}\right) $ is given
(\ref{1.81}). Therefore, using (\ref{1.80})-(\ref{1.82}), we can define the
2-d Radon transform $Rg$ of the function $g$ as 
\begin{equation}
\left( Rg\right) \left( x,x^{0}\right) =\left( Rg\right) \left( \alpha
,s\right) =\dint\limits_{\left\langle y,n\left( \alpha \right) \right\rangle
=s}g\left( y\right) d\sigma ,  \label{1.9}
\end{equation}

We are ready now to formulate Theorem 1, which is our main result.

\textbf{Theorem 1}. \emph{Suppose that the potential} $q(x)$ \emph{satisfies
conditions (\ref{1.1})-(\ref{1.3}). Let} $u_{sc}(x,x^{0},k)$ \emph{be the
function defined in (\ref{1.7}). Then for each pair of points }$x,x^{0}$, $%
x\neq x_{0}$ \emph{the asymptotic behavior of this function is} 
\begin{equation}
u_{sc}\left( x,x^{0},k\right) =\frac{i\exp (-ik|x-x^{0}|)}{8\pi |x-x^{0}|k}%
\left[ \dint\limits_{L\left( x,x_{0}\right) }q\left( \xi \right) d\sigma +O%
\Big(\frac{1}{k}\Big)\right] ,\>k\rightarrow \infty .  \label{1.90}
\end{equation}%
\emph{Hence, the asymptotic behavior of the function }$f(x,x^{0},k)$\emph{\
defined in (\ref{1.8}) is}%
\begin{equation}
f(x,x^{0},k)=\frac{1}{8\pi |x-x^{0}|k}\left[ (Rq)(x,x^{0})+O\Big(\frac{1}{k}%
\Big)\right] ,\>k\rightarrow \infty ;\forall x,x^{0}\in S_{a},x\neq x^{0},
\label{1.91}
\end{equation}%
\emph{for all }$a\in \left( -B,B\right) .$ \emph{Thus, for }$\left(
y,a\right) =\left( y_{1},y_{2},a\right) \in Q_{a},a\in \left( -B,B\right) $%
\emph{\ the reconstruction formula for the function }$q\left(
y_{1},y_{2},a\right) $\emph{\ is}%
\begin{equation}
q(y_{1},y_{2},a)=8\pi R^{-1}\{|x-x^{0}|\lim_{k\rightarrow \infty
}[kf(x,x^{0},k)]\}(y_{1},y_{2},a),\text{ }x,x^{0}\in S_{a}.  \label{1.10}
\end{equation}%
\emph{In (\ref{1.91}) and (\ref{1.10}) the operator }$(Rq)(x,x^{0})=(Rq)(%
\alpha ,s)$ \emph{of the 2-d Radon transform is defined as in (\ref{1.9})
via taking into account (\ref{1.80}) and (\ref{1.81}) and the same is true
for its inverse }$R^{-1}.$

\textbf{Remark 2. }The inversion formula (\ref{1.10}) follows immediately
from (\ref{1.82}), (\ref{1.9}), (\ref{1.91}) and the results of the book of
Natterer \cite{Nat}. Thus, we focus below on the proof of (\ref{1.90}),
since (\ref{1.91}) follows from (\ref{1.90}) immediately. It is well known
how to explicitly construct the operator $R^{-1},$ see, e.g. \cite{Nat}.
Hence, we are not doing this here for brevity.

\section{Proof of Theorem 1}

\label{sec:3}

We assume everywhere below that conditions of Theorem 1 are satisfied. To
prove Theorem 1, we consider first in subsection 3.1 the fundamental
solution of a hyperbolic PDE and formulate Lemma 1 about the $C^{2}-$%
smoothness of its regular part above the characteristic cone. It is known
how to establish the $C^{\infty }-$smoothness of the fundamental solution of
a hyperbolic equation with $C^{\infty }-$ coefficients, see, e.g. section
2.2 in the book of Romanov \cite{Rom}.\ However, we want to use here only
the $C^{4}-$smoothness of the potential $q\left( x\right) $ as in (\ref{1.2}%
). Hence, the proof of Lemma 1 is quite technical. For this reason, we
present that proof in section 4. We prove formula (\ref{1.90}) in subsection
3.2.

\subsection{The fundamental solution a hyperbolic equation}

\label{sec:3.1}

We now consider the following Cauchy problem%
\begin{eqnarray}
&&w_{tt}=\Delta _{x}w-q(x)w+4\pi \delta (x-x^{0},t),\quad (x,t)\in \mathbb{R}%
^{4},  \label{3.1} \\
&&w|_{t<0}=0.  \label{3.2}
\end{eqnarray}%
For an arbitrary $T>0$ denote 
\begin{equation*}
G(x^{0},T)=\{(x,t):\,0<|x-x^{0}|<t\leq T\}.
\end{equation*}%
For $t>|x-x^{0}|$ consider the ellipsoid $E(x,x^{0},t)$, 
\begin{equation*}
E\left( x,x^{0},t\right) =\left\{ \xi \in \mathbb{R}^{3}:\left\vert x-\xi
\right\vert +\left\vert x^{0}-\xi \right\vert =t\right\} .
\end{equation*}%
It was shown in \S 1 of Chapter 7 of the book of Lavrentiev,\ Romanov and
Shishatskii \cite{LRS} that the function $w\left( x,x^{0},t\right) =0$ for $%
t<\left\vert x-x^{0}\right\vert $ and it can be represented as%
\begin{equation}
w\left( x,x^{0},t\right) =w_{0}\left( x,x^{0},t\right) +\tilde{w}\left(
x,x^{0},t\right) H(t-|x-x^{0}|),  \label{3.3}
\end{equation}%
where $H(t)$ is the Heaviside function and functions $w_{0}$ and $\tilde{w}$
are defined by the following formulae: 
\begin{equation}
w_{0}\left( x,x^{0},t\right) =\frac{\delta \left( t-\left\vert
x-x^{0}\right\vert \right) }{\left\vert x-x^{0}\right\vert },  \label{3.4}
\end{equation}%
\begin{equation}
\tilde{w}\left( x,x^{0},t\right) =w_{1}\left( x,x^{0},t\right)
+\dsum\limits_{n=2}^{\infty }w_{n}\left( x,x^{0},t\right) ,  \label{3.5}
\end{equation}%
\begin{eqnarray}
w_{1}\left( x,x^{0},t\right) &=&-\frac{1}{4\pi (t^{2}-\rho ^{2})}%
\dint\limits_{E\left( x,x^{0},t\right) }r^{2}q(\xi )d\omega ,  \notag \\
&=&-\frac{1}{4\pi \rho }\int\limits_{(t-\rho )/2}^{(t+\rho
)/2}\int\limits_{0}^{2\pi }q(\xi )d\varphi dr,\quad \>t>\rho ,  \label{3.6}
\end{eqnarray}%
\begin{eqnarray}
w_{n}\left( x,x^{0},t\right) &=&-\frac{1}{4\pi }\int\limits_{\rho }^{t}\frac{%
1}{\tau ^{2}-\rho ^{2}}\left[ \dint\limits_{E(x,x^{0},\tau )}r^{3}q(\xi
)w_{n-1}(\xi ,x^{0},t-|x-\xi |)d\omega \right] d\tau  \notag \\
&=&-\frac{1}{4\pi \rho }\int\limits_{\rho }^{t}\left[ \dint%
\limits_{E(x,x^{0},\tau )}rq(\xi )w_{n-1}(\xi ,x^{0},t-\tau +r)d\varphi dr%
\right] d\tau ,  \label{3.7} \\
t &>&\rho ,\quad n\geq 2,  \notag
\end{eqnarray}%
where $\rho =|x-x^{0}|$, $r=|\xi -x^{0}|$, $d\omega =\sin \theta d\theta
d\varphi $ and $\xi $ is given by the formulae 
\begin{eqnarray*}
\xi &=&x^{0}+r\nu (\theta ,\varphi )A(\vartheta ,\psi ),\quad x=x^{0}+\rho
\nu (\vartheta ,\psi ), \\
\nu (\theta ,\varphi ) &=&(\sin \theta \cos \varphi ,\sin \theta \sin
\varphi ,\cos \theta ),\quad \\
A(\vartheta ,\psi ) &=&\left( 
\begin{array}{ccc}
-\cos \vartheta \cos \psi & -\cos \vartheta \sin \psi & \sin \vartheta \\ 
\sin \psi & -\cos \psi & 0 \\ 
\sin \vartheta \cos \psi & \sin \vartheta \sin \psi & \cos \vartheta%
\end{array}%
\right) .
\end{eqnarray*}%
Here $\rho $, $\vartheta \in \left[ 0,\pi \right] $ and $\psi \in \left[
0,2\pi \right) $ are spherical coordinates of the vector $x-x^{0}$ with the
center at $\left\{ x^{0}\right\} .$ Next, $r$, $\theta $ and $\varphi $ are
spherical coordinates of the vector $\xi -x^{0}$ with respect to the new
coordinates system $\xi _{1}^{\prime }$, $\xi _{2}^{\prime }$, $\xi
_{3}^{\prime }$. The center of this new system is $\left\{ x^{0}\right\} ,$
the axis $\xi _{3}^{\prime }$ passes through points $x^{0}$ and $x,$ and
axis $\xi _{1}^{\prime },\xi _{2}^{\prime }$ are orthogonal both to $\xi
_{3}^{\prime }$ and to each other and the axis $\xi _{1}^{\prime }$ lays in
the plane passing through the axis $\xi _{3}$ and $\xi _{3}^{\prime }$. The
orientation of the system $\xi _{1}^{\prime }$, $\xi _{2}^{\prime }$, $\xi
_{3}^{\prime }$ is the same as the orientation of the system $\xi _{1}$, $%
\xi _{2}$, $\xi _{3}$. Further, $\theta \in \left[ 0,\pi \right] $ is the
angle between the vector $\xi -x^{0}$ and the axis $\xi _{3}^{\prime }.$ In
the ellipsoid $E(x,x^{0},t)$ variables $r$ and $\theta $ are connected via 
\begin{equation}
r=\frac{t^{2}-\rho ^{2}}{2(t-\rho \cos \theta )},\quad \theta =\arccos \Big(%
\frac{2tr-t^{2}+\rho ^{2}}{2r\rho }\Big).  \label{3.9}
\end{equation}%
It was shown in \cite{LRS} that the series (\ref{3.5}) converges uniformly
in $\overline{G(x^{0},T)}=\{(x,t):|x-x^{0}|\leq t\leq T\}$\emph{\ }for any $%
T>0$ and, moreover, 
\begin{equation}
\lim_{t\rightarrow |x-x^{0}|^{+}}w_{1}(x,x^{0},t)=-\frac{1}{2|x-x^{0}|}%
\int\limits_{L(x,x^{0})}q(\xi )d\sigma ,  \label{3.10}
\end{equation}%
\begin{equation}
\lim_{t\rightarrow \left\vert x-x^{0}\right\vert ^{+}}w_{n}\left(
x,x^{0},t\right) =0,\quad n\geq 2.  \label{3.11}
\end{equation}%
Hence, 
\begin{equation}
\lim_{t\rightarrow \left\vert x-x^{0}\right\vert ^{+}}\tilde{w}\left(
x,x^{0},t\right) =-\frac{1}{2\left\vert x-x^{0}\right\vert }%
\int\limits_{L\left( x,x^{0}\right) }q\left( \xi \right) d\sigma =-\frac{1}{2%
}\int\limits_{0}^{1}q(x^{0}+z(x-x^{0}))dz.  \label{3.12}
\end{equation}

Lemma 1 and Corollary 1 guarantee a certain smoothness of the function $%
\tilde{w}.$

\textbf{Lemma 1.} \emph{For any }$T>0$ \emph{and for any} $x^{0}\in \mathbb{R%
}^{3}$\emph{\ functions }$\partial _{t}^{k}\tilde{w}(x,x^{0},t)\in C\left( 
\overline{G(x^{0},T)}\right) $ for $k=0,1,2.$

\textbf{Corollary 1.} \emph{The function }$\Delta _{x}\tilde{w}\in C\left( 
\overline{G(x^{0},T)}\right) .$

\textbf{Proof}. By (\ref{3.1}), (\ref{3.4}) and (\ref{3.5}) 
\begin{equation*}
\Delta _{x}\tilde{w}=\tilde{w}_{tt}+q\left( x\right) \tilde{w}\quad \text{for%
}\>\text{ }t>|x-x^{0}|.
\end{equation*}%
By\ Lemma 1 the right hand side of this equation belongs to $C\left( 
\overline{G(x^{0},T)}\right) .$ Hence, the assertion of this Corollary is
true. $\square $

\subsection{Proof of (\protect\ref{1.90})}

\label{sec:3.2}

First, we show that functions $\partial _{t}^{k}w(x,x^{0},t),k=0,1,2$ and $%
\Delta _{x}w(x,x^{0},t)$ decay exponentially as $t\rightarrow \infty $ and $%
x $ reminds in a bounded domain. To do this, we refer to Lemma 6 of Chapter
10 of the book of Vainberg \cite{V} as well as to Remark 3 after that lemma.
It follows from these results as well as from Lemma 1 and Corollary 1 that
for every $R>0$ and domain $D(x^{0},R)=\{x\in \mathbb{R}^{3}:|x-x^{0}|<R\}$
there exist numbers $C_{2}=C_{2}>0,c_{2}=c_{2}>0,t_{0}>0$ depending only on $%
q,x^{0},R$ such that for $k=0,1,2$ 
\begin{equation}
\left\vert \partial _{t}^{k}w\left( x,x^{0},t\right) \right\vert ,\left\vert
\Delta _{x}w\left( x,x^{0},t\right) \right\vert \leq C_{2}e^{-c_{2}t}\text{
for all }\>t\geq t_{0}\>\text{ and for all }\>x\in D(x^{0},R).  \label{3.21}
\end{equation}%
By (\ref{3.21}) we can apply Fourier transform with respect to $t$ to
functions $\partial _{t}^{k}w\left( x,x^{0},t\right) ,$ $\Delta _{x}w\left(
x,x^{0},t\right) $. Let 
\begin{equation*}
v\left( x,x^{0},k\right) =\frac{1}{4\pi }\int\limits_{0}^{\infty }w\left(
x,x^{0},t\right) e^{-ikt}dt,\text{ }\forall x,x^{0}\in \mathbb{R}^{3},x\neq
x^{0},\forall k\in \mathbb{R}.
\end{equation*}%
Using again the same results of references \cite{GT,V1,V} as ones cited in
section 2, we obtain that 
\begin{equation}
u\left( x,x^{0},k\right) =v\left( x,x^{0},k\right) ,\forall x,x^{0}\in 
\mathbb{R}^{3},x\neq x^{0},\forall k\in \mathbb{R}.  \label{3.22}
\end{equation}

Consider now the asymptotic behavior of the function $u_{sc}\left(
x,x^{0},k\right) =u\left( x,x^{0},k\right) -u_{0}\left( x,x^{0},k\right) $
as $k\rightarrow \infty .$ Clearly%
\begin{equation}
u_{sc}\left( x,x^{0},k\right) =\frac{1}{4\pi }\int\limits_{\left\vert
x-x^{0}\right\vert }^{\infty }\tilde{w}\left( x,x^{0},t\right) e^{-ikt}dt,%
\text{ }\forall x,x^{0}\in \mathbb{R}^{3},x\neq x^{0},\forall k\in \mathbb{R}%
.  \label{3.23}
\end{equation}%
Using (\ref{3.10}), Lemma 1 and integration by parts, we obtain%
\begin{equation*}
\int\limits_{\left\vert x-x^{0}\right\vert }^{\infty }\tilde{w}\left(
x,x^{0},t\right) e^{-ikt}dt=
\end{equation*}%
\begin{equation*}
-\frac{i\exp \left( -ik\left\vert x-x^{0}\right\vert \right) }{k}\left[ -%
\frac{1}{8\pi \left\vert x-x^{0}\right\vert }\int\limits_{L\left(
x,x^{0}\right) }q\left( \xi \right) d\sigma +\frac{i}{k}\partial _{t}\tilde{w%
}\left( x,x^{0},\left\vert x-x^{0}\right\vert ^{+}\right) \right] 
\end{equation*}%
\begin{equation*}
-\frac{1}{k^{2}}\dint\limits_{\left\vert x-x^{0}\right\vert }^{\infty
}\partial _{t}^{2}\tilde{w}\left( x,x^{0},t\right) e^{-ikt}dt.
\end{equation*}

Hence, by (\ref{3.23}) the asymptotic behaviour of the function $%
u_{sc}\left( x,x^{0},k\right) $ is%
\begin{equation}
u_{sc}\left( x,x^{0},k\right) =\frac{i\exp \left( ik\left\vert
x-x^{0}\right\vert \right) }{8\pi k\left\vert x-x^{0}\right\vert }\left[
\int\limits_{L\left( x,x^{0}\right) }q\left( \xi \right) d\sigma +O\left( 
\frac{1}{k}\right) \right] ,\text{ }k\rightarrow \infty .  \label{3.24}
\end{equation}%
Since formula (\ref{3.24}) coincides with formula (\ref{1.90}), then the
proof of (\ref{1.90}) is complete. Next, since by Remark 2 the validity of
Theorem 1 follows from (\ref{1.90}), then the proof of this theorem is
complete as well. $\square $

\section{Proof of Lemma 1}

\label{sec:4}

Denote 
\begin{equation}
q_{0}=\left\Vert q\right\Vert _{C\left( \overline{\Omega }\right)
},q_{2}=\left\Vert q\right\Vert _{C^{2}\left( \overline{\Omega }\right)
},q_{4}=\left\Vert q\right\Vert _{C^{4}\left( \overline{\Omega }\right) }.
\label{3.0}
\end{equation}%
It follows from (\ref{3.10}) and (\ref{3.12}) that the function $%
w_{1}(x,x^{0},t)\in C\left( \overline{G(x^{0},T)}\right) .$ First, we prove
that the function $w_{1}$ given by formula (\ref{3.6}) has the first
derivative $\partial _{t}w_{1}$ which belongs to $C\left( \overline{%
G(x^{0},T)}\right) $. Change variables $r\Leftrightarrow z$ in the second
integral (\ref{3.6}) as 
\begin{equation}
r=\frac{1}{2}(t-\rho +2z\rho ).  \label{3.120}
\end{equation}%
Then 
\begin{equation}
w_{1}(x,x^{0},t)=-\frac{1}{4\pi }\int\limits_{0}^{1}\int\limits_{0}^{2\pi
}q(\xi )d\varphi dz,  \label{3.13}
\end{equation}%
where $\xi \in E(x,x^{0},t)$ is 
\begin{equation}
\xi =x^{0}+\frac{1}{2}(t-\rho +2z\rho )\nu (\theta ,\varphi )A(\vartheta
,\psi ),\quad \theta =\arccos \Big(\frac{\rho -t+2zt}{t-\rho +2z\rho }\Big).
\label{3.14}
\end{equation}%
Using (\ref{3.13}), we obtain 
\begin{equation}
\partial _{t}w_{1}(x,x^{0},t)=-\frac{1}{4\pi }\int\limits_{0}^{1}\int%
\limits_{0}^{2\pi }\nabla q(\xi )\cdot \partial _{t}\xi d\varphi dz.
\label{3.140}
\end{equation}%
Using (\ref{3.14}), we calculate $\partial _{t}\xi $ as 
\begin{eqnarray}
\partial _{t}\xi &=&\frac{1}{2}\nu (\theta ,\varphi )A(\vartheta ,\psi )+%
\frac{1}{2}(t-\rho +2z\rho )\nu _{\theta }(\theta ,\varphi )A(\vartheta
,\psi )\partial _{t}\theta  \notag \\
&=&\frac{1}{2}\nu (\theta ,\varphi )A(\vartheta ,\psi )+\frac{2\rho z(1-z)}{%
(t-\rho +2z\rho )\sin \theta }\nu _{\theta }(\theta ,\varphi )A(\vartheta
,\psi ),  \label{3.15}
\end{eqnarray}%
where $\nu _{\theta }(\theta ,\varphi )=\partial _{\theta }\nu (\theta
,\varphi )$. In (\ref{3.15}) we have used the following formula: 
\begin{equation}
\partial _{t}\theta =\frac{4\rho z(1-z)}{(t-\rho +2z\rho )^{2}\sin \theta }.
\label{3.30}
\end{equation}%
Indeed, by (\ref{3.14})%
\begin{equation*}
\cos \theta =\frac{\rho -t+2zt}{t-\rho +2z\rho }.
\end{equation*}%
Hence, 
\begin{equation*}
-\sin \theta \partial _{t}\theta =\frac{(2z-1)(t-\rho +2z\rho )-(\rho -t+2zt)%
}{(t-\rho +2z\rho )^{2}}=\frac{-4z\rho (1-z)}{(t-\rho +2z\rho )^{2}},
\end{equation*}%
which proves (\ref{3.30}).

Hence, 
\begin{equation}
\nabla q(\xi )\cdot \partial _{t}\xi =Q_{1}\cos \varphi +Q_{2}\sin \varphi
+Q_{3}(2z-1),  \label{3.31}
\end{equation}%
where functions $Q_{j}=Q_{j}(\xi ,t,z,\theta ,\rho ,\vartheta ,\psi )$, $%
j=1,2,3$ are:\textbf{\ } 
\begin{eqnarray*}
Q_{1} &=&\Big[\frac{2\rho z(1-z)\cot \theta }{(t-\rho +2z\rho )}+\frac{1}{2}%
\sin \theta \Big]\sum_{k=1}^{3}(\partial _{\xi _{k}}q)a_{1k}, \\
Q_{2} &=&\Big[\frac{2\rho z(1-z)\cot \theta }{(t-\rho +2z\rho )}+\frac{1}{2}%
\sin \theta \Big]\sum_{k=1}^{3}(\partial _{\xi _{k}}q)a_{2k}, \\
Q_{3} &=&\frac{1}{2}\sum_{k=1}^{3}(\partial _{\xi _{k}}q)a_{3k},
\end{eqnarray*}%
where\textbf{\ }$a_{jk}(\vartheta ,\psi )$ are elements of the matrix $A$.
We now calculate integrals over the ellipsoid $E(x,x^{0},t)$ separately for
the each term $Q_{j}$, $j=1,2,3$. Using the integration by parts with
respect to $\varphi $, we obtain 
\begin{eqnarray}
I_{1}(x,x^{0},t) &=&-\frac{1}{4\pi }\int\limits_{0}^{1}\int\limits_{0}^{2\pi
}Q_{1}\cos \varphi d\varphi dz=\frac{1}{4\pi }\int\limits_{0}^{1}\int%
\limits_{0}^{2\pi }(\partial _{\varphi }Q_{1})\sin \varphi d\varphi dz 
\notag \\
&=&\frac{1}{4\pi }\int\limits_{0}^{1}\int\limits_{0}^{2\pi
}\sum_{m=1}^{3}(\partial _{\xi _{m}}Q_{1})(\partial _{\varphi }\xi
_{m})a_{1k}\sin \varphi d\varphi dz  \notag \\
&=&\frac{1}{4\pi }\int\limits_{0}^{1}\int\limits_{0}^{2\pi }\Big[\rho
z(1-z)\cos \theta +\frac{(t-\rho +2z\rho )}{4}\sin ^{2}\theta \Big]
\label{3.32} \\
&&\times \sum_{k,m=1}^{3}(\partial _{\xi _{k},\xi _{m}}^{2}q)a_{1k}\sin
\varphi (a_{2m}\cos \varphi -a_{1m}\sin \varphi )d\varphi dz.  \notag
\end{eqnarray}%
Similarly, 
\begin{eqnarray}
I_{2}(x,x^{0},t) &=&-\frac{1}{4\pi }\int\limits_{0}^{1}\int\limits_{0}^{2\pi
}Q_{2}\sin \varphi d\varphi dz  \notag \\
&=&-\frac{1}{4\pi }\int\limits_{0}^{1}\int\limits_{0}^{2\pi }\Big[\rho
z(1-z)\cos \theta +\frac{(t-\rho +2z\rho )}{4}\sin ^{2}\theta \Big]
\label{3.33} \\
&&\times \sum_{k,m=1}^{3}(\partial _{\xi _{k},\xi _{m}}^{2}q)a_{2k}\cos
\varphi (a_{2m}\cos \varphi -a_{1m}\sin \varphi )d\varphi dz.  \notag
\end{eqnarray}

Consider now the integral with $Q_{3}$. Since $(2z-1)dz=d(z^{2}-z)$, then
the integration by parts with respect to $z$ leads to 
\begin{eqnarray}
I_{3}(x,x^{0},t) &=&-\frac{1}{4\pi }\int\limits_{0}^{1}\int\limits_{0}^{2\pi
}Q_{3}(2z-1)d\varphi dz  \notag \\
&=&-\frac{\rho }{8\pi }\int\limits_{0}^{1}\int\limits_{0}^{2\pi
}z(1-z)\sum_{k,m=1}^{3}(\partial _{\xi _{k},\xi _{m}}^{2}q)a_{3k}
\label{3.34} \\
&&\times \lbrack \sin \theta (a_{1m}\cos \varphi +a_{2m}\sin \varphi
)+a_{3m}\cos \theta ]d\varphi dz.  \notag
\end{eqnarray}%
Hence, using (\ref{3.140}), (\ref{3.31})-(\ref{3.34}), we obtain 
\begin{eqnarray}
\partial _{t}w_{1}(x,x^{0},t) &=&-\frac{\rho }{8\pi }\int\limits_{0}^{1}\int%
\limits_{0}^{2\pi }\sum_{k,m=1}^{3}(\partial _{\xi _{k},\xi _{m}}^{2}q)\Big\{%
\Big[z(1-z)\cos \theta +\frac{(t-\rho +2z\rho )}{4}\sin ^{2}\theta \Big] 
\notag \\
&&\times \lbrack b_{0km}+b_{1km}\cos (2\varphi )+b_{2km}\sin (2\varphi )]
\label{3.35} \\
&&+z(1-z)a_{3k}[\sin \theta (a_{1m}\cos \varphi +a_{2m}\sin \varphi
)+a_{3m}\cos \theta ]\Big\}d\varphi dz,  \notag
\end{eqnarray}%
where 
\begin{equation}
b_{0km}=a_{1k}a_{1m}+a_{2k}a_{2m},\quad
b_{1km}=a_{2k}a_{2m}-a_{1k}a_{1m},\quad b_{2km}=-a_{2k}a_{1m}-a_{1k}a_{2m}.
\label{3.36}
\end{equation}%
Thus, it follows from (\ref{3.35}) and (\ref{3.36}) that the function $%
\partial _{t}w_{1}\in C\left( \overline{G(x^{0},T)}\right) $ and the
following estimate holds 
\begin{equation*}
|\partial _{t}w_{1}(x,x^{0},t)|\leq Cq_{2}\text{ in }\emph{\ }\overline{%
G(x^{0},T)},
\end{equation*}%
where the number $q_{2}$ is defined in (\ref{3.0}). Here and below $C=C(T)$
is a positive constant depending only on $T$. Since $\cos \theta \rightarrow
1$ and $\xi \rightarrow x^{0}+z(x-x^{0}),$ as $t\rightarrow |x-x^{0}|^{+},$
then 
\begin{eqnarray*}
\lim_{t\rightarrow |x-x^{0}|^{+}}\partial _{t}w_{1}(x,x^{0},t) &=&-\frac{%
\rho }{8}\sum_{k,m=1}^{3}\int\limits_{0}^{1}{\ z(1-z)}(\partial _{\xi
_{k},\xi _{m}}^{2}q)(x^{0}+z(x-x^{0})(b_{0km}+a_{3k}a_{3m})dz, \\
&=&-\frac{\rho }{8}\int\limits_{0}^{1}{\ z(1-z)}(\Delta
q)(x^{0}+z(x-x^{0})dz.
\end{eqnarray*}%
In the latter equality we use the fact that 
\begin{equation*}
b_{0km}+a_{3k}a_{3m}=\sum_{j=1}^{3}a_{jk}a_{jm}=\delta _{km},
\end{equation*}%
where $\delta _{km}$ is the Kronecker delta, since the matrix $A$ is
orthogonal.

Consider now functions $w_{n}(x,x^{0},t)$, $n\geq 2$ given by (\ref{3.7}).
We show that each function $w_{n}(x,x^{0},t)\in C\left( \overline{G\left(
x^{0},T\right) }\right) $ and the following estimate holds 
\begin{equation}
|w_{n}(x,x^{0},t)|\leq \frac{q_{0}^{n}T^{n-1}(t-\rho )^{n-1}}{4(n-1)!},\quad
n\geq 2,  \label{3.37}
\end{equation}%
where the number $q_{0}$ is defined in (\ref{3.0}). First, we observe that
the same estimate is valid for $n=1$. Indeed, taking into account that by (%
\ref{3.6}) and (\ref{3.9}) $r\leq (\tau +\rho )/2\leq T$ for $\xi \in
E(x,x^{0},\tau )$, we obtain 
\begin{eqnarray}
|w_{1}(x,x^{0},t)| &\leq &\frac{1}{4\pi }\int\limits_{0}^{1}\int%
\limits_{0}^{2\pi }q_{0}d\varphi dz=\frac{q_{0}}{2},  \notag \\
|w_{2}(x,x^{0},t)| &\leq &\frac{1}{4\pi }\int\limits_{\rho }^{t}\left(
\int\limits_{0}^{1}\int\limits_{0}^{2\pi }Tq_{0}\frac{q_{0}}{2}d\varphi
dz\right) d\tau =\frac{Tq_{0}^{2}}{4}(t-\rho ),  \label{3.38} \\
|w_{3}(x,x^{0},t)| &\leq &\frac{1}{4\pi }\int\limits_{\rho }^{t}\left(
\int\limits_{0}^{1}\int\limits_{0}^{2\pi }Tq_{0}\frac{Tq_{0}^{2}}{4}\Big(t-%
\frac{\tau +\rho }{2}\Big)d\varphi dz\right) d\tau \leq \frac{T^{2}q_{0}^{3}%
}{4}\frac{(t-\rho )^{2}}{2!}.  \notag
\end{eqnarray}%
Continuing this way, we obtain estimate (\ref{3.37}) by the method of
mathematical induction.

Differentiating formula (\ref{3.7}) with respect to $t$, we obtain 
\begin{equation}
\partial _{t}w_{n}(x,x^{0},t)=-\frac{1}{4\pi \rho }\dint%
\limits_{E(x,x^{0},t)}rq(\xi )w_{n-1}(\xi ,x^{0},r)d\varphi dr  \label{3.39}
\end{equation}%
\begin{equation*}
-\frac{1}{4\pi \rho }\int\limits_{\rho }^{t}\left[ \dint\limits_{E(x,x^{0},%
\tau )}rq(\xi )\partial _{t}w_{n-1}(\xi ,x^{0},t-\tau +r)d\varphi dr\right]
d\tau ,\quad n\geq 2.
\end{equation*}%
If $n=2$, then the latter formula implies that the function $\partial
_{t}w_{2}(x,x^{0},t)$ is continuous in $G(x^{0},T)$ and 
\begin{eqnarray*}
\lim_{t\rightarrow |x-x^{0}|^{+}}\partial _{t}w_{2}(x,x^{0},t) &=&-\frac{1}{2%
}\int%
\limits_{0}^{1}zq(x^{0}+z(x-x^{0})w_{1}(x^{0}+z(x-x^{0}),x^{0},|x-x^{0}|)dz
\\
&=&\frac{1}{4}\int\limits_{0}^{1}zq(x^{0}+z(x-x^{0})\int%
\limits_{0}^{1}q(x^{0}+z_{1}z(x-x^{0})dz_{1}dz.
\end{eqnarray*}%
Hence, $\partial _{t}w_{2}\in C\left( \overline{G\left( x^{0},T\right) }%
\right) $. Also, we obtain the following estimate from the above discussion 
\begin{equation}
|\partial _{t}w_{2}(x,x^{0},t)|\leq \frac{Tq_{0}}{2}\Big(\frac{q_{0}}{2}%
+CTq_{2}\Big)\leq \frac{Tq_{0}}{4}C_{1}q_{2},\text{ in }G(x^{0},T),
\label{3.40}
\end{equation}%
where $C_{1}=1+2TC$. If $n\geq 3,$ then the first integral in formula (\ref%
{3.39}) vanishes, which follows immediately from estimate (\ref{3.37}).
Indeed, by this estimate $w_{n-1}(\xi ,x^{0},|\xi -x^{0}|)=0$ for $n\geq 3$.
As to the second integral in (\ref{3.39}), it obviously tends to zero as $%
t\rightarrow |x-x^{0}|^{+}$.

Thus, by (\ref{3.39}) 
\begin{equation}
\lim_{t\rightarrow |x-x^{0}|^{+}}\partial _{t}w_{n}(x,x^{0},t)=0,\quad n\geq
3.  \label{3.41}
\end{equation}

Continuing estimates (\ref{3.40}) for $n=3,4,...$ and taking into account (%
\ref{3.39}) and (\ref{3.41}), we obtain by the method of mathematical
induction that all functions $\partial _{t}w_{n}\in C\left( \overline{%
G\left( x^{0},T\right) }\right) ,n\geq 1,$ and the following estimates are
valid

\begin{equation*}
|\partial _{t}w_{n}(x,x^{0},t)|\leq \frac{C_{1}q_{2}T^{(n-1)}q_{0}^{n-1}(t-%
\rho )^{n-2}}{4(n-2)!},\quad n\geq 3.
\end{equation*}

Hence, the series%
\begin{equation*}
\sum_{n=1}^{\infty }\partial _{t}w_{n}(x,x^{0},t)
\end{equation*}%
converges uniformly in $\overline{G(x^{0},T)}$ and its sum $\partial _{t}%
\tilde{w}\in C\left( \overline{G\left( x^{0},T\right) }\right) .$ Thus, we
have proved the assertion of Lemma 1 for functions $\tilde{w}$ and $\partial
_{t}\tilde{w}$.

Now we prove the assertion of this lemma for the second derivative $\partial
_{t}^{2}\tilde{w}(x,x^{0},t)$. Since the proof of this fact is quite similar
to the previous one, we omit some details for brevity. We represent the
formula for the first derivative of the function $w_{1}$ in the form: 
\begin{equation*}
\partial _{t}w_{1}(x,x^{0},t)=-\frac{\rho }{8\pi }\int\limits_{0}^{1}\int%
\limits_{0}^{2\pi }Sd\varphi dz,
\end{equation*}%
where 
\begin{equation*}
S=S_{1}\cos \theta +S_{2}\sin \theta +S_{3}\sin ^{2}\theta ,
\end{equation*}%
where functions $S_{j}=S_{j}(\xi ,t,z,\varphi ,\rho ,\vartheta ,\psi )$ are
defined by: 
\begin{eqnarray*}
S_{1} &=&\rho z(1-z)\sum_{k,m=1}^{3}(\partial _{\xi _{k},\xi
_{m}}^{2}q)[\delta _{km}+b_{1km}\cos (2\varphi )+b_{2km}\sin (2\varphi )], \\
S_{2} &=&\rho z(1-z)\sum_{k,m=1}^{3}(\partial _{\xi _{k},\xi
_{m}}^{2}q)a_{3k}[(a_{1m}\cos \varphi +a_{2m}\sin \varphi ), \\
S_{3} &=&\frac{(t-\rho +2z\rho )}{4}\sum_{k,m=1}^{3}(\partial _{\xi _{k},\xi
_{m}}^{2}q)[b_{0km}+b_{1km}\cos (2\varphi )+b_{2km}\sin (2\varphi )].
\end{eqnarray*}%
Hence, 
\begin{equation*}
\partial _{t}^{2}w_{1}(x,x^{0},t)=-\frac{\rho }{8\pi }\int\limits_{0}^{1}%
\int\limits_{0}^{2\pi }\left[ \nabla _{\xi }S\cdot \partial _{t}\xi
+(\partial _{t}S_{3})\sin ^{2}\theta +\left( S_{2}\cos \theta -S_{1}\sin
\theta +S_{3}\sin (2\theta )\right) \partial _{t}\theta \right] d\varphi dz.
\end{equation*}%
We have 
\begin{eqnarray*}
\nabla S\cdot \partial _{t}\xi &=&\sum_{i=1}^{3}\Big[\frac{2\rho z(1-z)\coth
\theta }{(t-\rho +2z\rho )}+\frac{1}{2}\sin \theta \Big](\partial _{\xi
_{i}}S) \\
&&\times \lbrack a_{1i}\cos \varphi +a_{2i}\sin \varphi ]+\frac{2z-1}{2}%
\sum_{i=1}^{3}(\partial _{\xi _{i}}S)a_{3i}.
\end{eqnarray*}%
Hence, we represent $\partial _{t}^{2}w_{1}(x,x^{0},t)$ as 
\begin{equation}
\partial _{t}^{2}w_{1}(x,x^{0},t)=-\frac{\rho }{8\pi }\int\limits_{0}^{1}%
\int\limits_{0}^{2\pi }[R_{1}+R_{2}]d\varphi dz,  \label{3.42}
\end{equation}%
where 
\begin{eqnarray*}
R_{1} &=&\Big[\frac{2\rho z(1-z)\cos \theta }{(t-\rho +2z\rho )}%
\sum_{i=1}^{3}(\partial _{\xi _{i}}(S_{2}+S_{3}\sin \theta ))+\frac{1}{2}%
\sin \theta \sum_{i=1}^{3}(\partial _{\xi _{i}}S)\Big]\lbrack a_{1i}\cos
\varphi +a_{2i}\sin \varphi ] \\
&&+\frac{2z-1}{2}\sum_{i=1}^{3}(\partial _{\xi _{i}}S)a_{3i}+\frac{1}{4}%
\sum_{k,m=1}^{3}(\partial _{\xi _{k},\xi _{m}}^{2}q)[b_{0km}+b_{1km}\cos
(2\varphi )+b_{2km}\sin (2\varphi )]\sin ^{2}\theta \\
&&+\frac{4\rho z(1-z)}{(t-\rho +2z\rho )^{2}}(-S_{1}+2S_{3}\cos \theta ), \\
R_{2} &=&\frac{2\rho z(1-z)\cos \theta }{(t-\rho +2z\rho )\sin \theta }\Big[%
\cos \theta \sum_{i=1}^{3}(\partial _{\xi _{i}}S_{1})[a_{1i}\cos \varphi
+a_{2i}\sin \varphi ]+\frac{2S_{2}}{(t-\rho +2z\rho )}\Big].
\end{eqnarray*}

The term $R_{1}$ is obviously bounded.\ However, $R_{2}$ is unbounded.
Hence, we represent $R_{2}$ as 
\begin{equation*}
R_{2}=\frac{z(1-z)^{2}\cos \theta }{(t-\rho +2\rho z)\sin \theta }%
\sum_{s=1}^{3}[M_{s}\cos (s\varphi )+N_{s}\sin (s\varphi )],
\end{equation*}

where $M_{s}=M_{s}(\xi ,t,z,\rho ,\vartheta ,\psi )$ and $N_{s}=N_{s}(\xi
,t,z,\rho ,\vartheta ,\psi )$ are defined by: 
\begin{eqnarray*}
M_{1} &=&\rho z\cos \theta \sum_{k,m,i=1}^{3}(\partial _{\xi _{k},\xi
_{m},\xi _{i}}^{3}q)(2\delta _{km}a_{1i}+b_{1km}a_{1i}+b_{2km}a_{2i}) \\
&&+\frac{2\rho z}{(t-\rho +2\rho z)}\sum_{k,m=1}^{3}(\partial _{\xi _{k},\xi
_{m}}^{2}q)a_{3k}a_{1m},
\end{eqnarray*}%
\begin{eqnarray*}
N_{1} &=&\rho z\cos \theta \sum_{k,m,i=1}^{3}(\partial _{\xi _{k},\xi
_{m},\xi _{i}}^{3}q)(2\delta _{km}a_{2i}+b_{2km}a_{1i}-b_{1km}a_{2i}) \\
&&+\frac{2\rho z}{(t-\rho +2\rho z)}\sum_{k,m=1}^{3}(\partial _{\xi _{k},\xi
_{m}}^{2}q)a_{3k}a_{2m},
\end{eqnarray*}
\begin{equation*}
M_{2}=N_{2}=0,
\end{equation*}
\begin{eqnarray*}
M_{3} &=&\rho z\cos \theta \sum_{k,m,i=1}^{3}(\partial _{\xi _{k},\xi
_{m},\xi _{i}}^{3}q)(b_{1km}a_{1i}-b_{2km}a_{2i}) \\
N_{3} &=&\rho z\cos \theta \sum_{k,m,i=1}^{3}(\partial _{\xi _{k},\xi
_{m},\xi _{i}}^{3}q)(b_{2km}a_{1i}+b_{1km}a_{2i}).
\end{eqnarray*}

By (\ref{3.42}) 
\begin{equation*}
\partial _{t}^{2}w_{1}(x,x^{0},t)=J_{1}(x,x^{0},t)+J_{2}(x,x^{0},t),
\end{equation*}%
\begin{equation*}
J_{j}(x,x^{0},t)=-\frac{\rho }{8\pi }\int\limits_{0}^{1}\int\limits_{0}^{2%
\pi }R_{j}d\varphi dz,\quad j=1,2.
\end{equation*}%
The function $J_{1}(x,x^{0},t)$ is obviously continuous and bounded for $%
(x,t)\in \overline{G(x^{0},T)}$. We show now that the function $%
J_{2}(x,x^{0},t)$ has the same properties. We have 
\begin{eqnarray*}
J_{2}(x,x^{0},t) &=&-\frac{\rho }{8\pi }\int\limits_{0}^{1}\int%
\limits_{0}^{2\pi }\frac{z(1-z)^{2}\cot \theta }{(t-\rho +2\rho z)}%
\sum_{s=1}^{3}\Big[M_{s}\cos (s\varphi )+N_{s}\sin (s\varphi )]d\varphi \Big]%
dz \\
&=&\frac{\rho }{8\pi }\int\limits_{0}^{1}\int\limits_{0}^{2\pi }\frac{%
z(1-z)^{2}\cot \theta }{(t-\rho +2\rho z)}\sum_{s,j=1}^{3}\Big[\frac{1}{s}%
\sin (s\varphi )(\partial _{\xi _{j}}M_{s})-\frac{1}{s}\cos (s\varphi
)(\partial _{\xi _{j}}N_{s})\Big](\partial _{\varphi }\xi _{j})d\varphi dz.
\end{eqnarray*}

Hence, 
\begin{eqnarray*}
J_{2}(x,x^{0},t) &=&\frac{\rho }{16\pi }\int\limits_{0}^{1}\int%
\limits_{0}^{2\pi }z(1-z)^{2}\Big[\sum_{s,j=1}^{3}\frac{1}{s}(\sin (s\varphi
)\partial _{\xi _{j}}M_{s}-\cos (s\varphi )\partial _{\xi _{j}}N_{s}) \\
&&\times (a_{2j}\cos \varphi -a_{1j}\sin \varphi )\Big]d\varphi dz.
\end{eqnarray*}

Since functions $\partial _{\xi _{j}}M_{s}$ and $\partial _{\xi _{j}}N_{s}$
are bounded for $(x,t)\in \overline{G(x^{0},T)},$ the function $%
J_{2}(x,x^{0},t)$ is also bounded and continuous in $G(x^{0},T).$ Thus, the
functions $J_{1},J_{2},\partial _{t}^{2}w_{1}$ are bounded and continuos in $%
\overline{G(x^{0},T)}$. Simple estimates lead to 
\begin{equation}
|\partial _{t}^{2}w_{1}(x,x^{0},t)|\leq Cq_{4}.  \label{300}
\end{equation}

Moreover, 
\begin{eqnarray*}
\lim_{t\rightarrow |x-x^{0}|^{+}}J_{1}(x,x^{0},t) &=&-\frac{1}{8}%
\int\limits_{0}^{1}\Big[\rho z(1-z)\sum_{k,m,i=1}^{3}(\partial _{\xi
_{k},\xi _{m},\xi _{i}}^{3}q) \\
&&\times \Big[a_{3k}(a_{1m}a_{1i}+a_{2m}a_{2i})+(2z-1)a_{3i}\delta _{km} \\
&&-2(1-z)^{2}\sum_{k,m=1}^{3}(\partial _{\xi _{k},\xi _{m}}^{3}q)a_{3k}a_{3m}%
\Big]dz,
\end{eqnarray*}%
\begin{eqnarray*}
\lim_{t\rightarrow |x-x^{0}|^{+}}J_{2}(x,x^{0},t) &=&-\frac{\rho }{32\pi }%
\int\limits_{0}^{1}\int\limits_{0}^{2\pi }z(1-z)^{2}\Big[\Big(\rho
z\sum_{k,m,i,j=1}^{3}(\partial _{\xi _{k},\xi _{m},\xi _{i},\xi _{j}}^{4}q)
\\
&&\times \lbrack a_{1j}(2a_{1i}\delta
_{km}+b_{1km}a_{1i}+b_{2km}a_{2i})+a_{2j}(a_{2i}\delta
_{km}-b_{1km}a_{2i}+b_{2km}a_{1i})]\Big) \\
&&+\sum_{k,m,j=1}^{3}(\partial _{\xi _{k},\xi _{m},\xi
_{j}}^{3}q)a_{3k}(a_{1m}a_{1j}+a_{2m}a_{2j})\Big]d\varphi dz.
\end{eqnarray*}

Hence, there exists the limit $\lim_{t\rightarrow |x-x^{0}|^{+}}\partial
_{t}^{2}w_{1}(x,x^{0},t)$ and it is continuous with respect to $x.$ Thus, we
have proven that the function $\partial _{t}^{2}w_{1}\in C\left( \overline{%
G(x^{0},T)}\right) .$

Consider now functions $\partial _{t}^{2}w_{n}$ for $n\geq 2$.
Differentiating formula (\ref{3.39}) with respect to $t$, we obtain 
\begin{equation*}
\partial _{t}^{2}w_{n}(x,x^{0},t)=-\frac{1}{4\pi \rho }\Bigg(\rho \partial
_{t}\int\limits_{0}^{1}\int\limits_{0}^{2\pi }rq(\xi )w_{n-1}(\xi
,x^{0},r)d\varphi dz
\end{equation*}%
\begin{equation}
+\dint\limits_{E(x,x^{0},t)}rq(\xi )\partial _{t}w_{n-1}(\xi
,x^{0},r)d\varphi dr\Big)  \label{3.20}
\end{equation}%
\begin{equation*}
+\int\limits_{\rho }^{t}d\tau \dint\limits_{E(x,x^{0},\tau )}rq(\xi
)\partial _{t}^{2}w_{n-1}(\xi ,x^{0},t-\tau +r)d\varphi dr\Bigg),\quad n\geq
2,
\end{equation*}%
where by (\ref{3.120}) $r=|\xi -x^{0}|=(t-\rho +2z\rho )/2$. Note that the
first integral in (\ref{3.20}) vanishes for $n>2$ since $%
w_{n-1}(x,x^{0},|x-x^{0}|)$ vanishes for for $n>2$. Hence, we should
calculate the derivative of this integral only for $n=2.$

Denote 
\begin{equation*}
Q(x,x^{0})=q(x)w_{1}(x,x^{0},|x-x^{0}|)=-\frac{q(x)}{2}\int%
\limits_{0}^{1}q(x^{0}+z(x-x^{0}))dz.
\end{equation*}%
The function $Q(x,x^{0})\in C^{4}(\mathbb{R}^{3}\times \mathbb{R}^{3})$. We
have 
\begin{equation*}
\partial _{t}\int\limits_{0}^{1}\int\limits_{0}^{2\pi }rq(\xi )w_{1}(\xi
,x^{0},r)d\varphi dz=\frac{1}{2}\partial
_{t}\int\limits_{0}^{1}\int\limits_{0}^{2\pi }(t-\rho +2z\rho )Q(\xi
,x^{0})dz
\end{equation*}%
\begin{equation}
=\frac{1}{2}\int\limits_{0}^{1}\int\limits_{0}^{2\pi }Q(\xi ,x^{0})d\varphi
dz+\frac{1}{2}\int\limits_{0}^{1}\int\limits_{0}^{2\pi }(t-\rho +2z\rho )%
\big(\nabla _{\xi }Q(\xi ,x^{0})\cdot \partial _{t}\xi \big)d\varphi dz.
\label{3.43}
\end{equation}%
The first of the integrals in the second line of (\ref{3.43}) is a bounded
and continuous function in $\overline{G(x^{0},T)}$ and 
\begin{equation*}
\lim_{t\rightarrow |x-x^{0}|^{+}}\frac{1}{2}\int\limits_{0}^{1}\int%
\limits_{0}^{2\pi }Q(\xi ,x^{0})d\varphi dz=\pi
\int\limits_{0}^{1}Q(x+z(x-x^{0}),x^{0})dz.
\end{equation*}%
The second integral in the second line of (\ref{3.43}) can be evaluated in
the same way as we have done above for the case of the first $t-$derivative.
More precisely, we use 
\begin{equation*}
\nabla Q(\xi ,x^{0})\cdot \partial _{t}\xi =\overline{Q}_{1}\cos \varphi +%
\overline{Q}_{2}\sin \varphi +\overline{Q}_{3}(2z-1),
\end{equation*}%
where functions $\overline{Q}_{j}=\overline{Q}_{j}(\xi ,x^{0},t,z,\theta
,\rho ,\vartheta ,\psi )$, $j=1,2,3$, are: 
\begin{eqnarray*}
\overline{Q}_{1} &=&\Big[\frac{2\rho z(1-z)\cot \theta }{(t-\rho +2z\rho )}+%
\frac{1}{2}\sin \theta \Big]\sum_{k=1}^{3}(\partial _{\xi _{k}}Q)a_{1k}, \\
\overline{Q}_{2} &=&\Big[\frac{2\rho z(1-z)\cot \theta }{(t-\rho +2z\rho )}+%
\frac{1}{2}\sin \theta \Big]\sum_{k=1}^{3}(\partial _{\xi _{k}}Q)a_{2k}, \\
\overline{Q}_{3} &=&\frac{1}{2}\sum_{k=1}^{3}(\partial _{\xi _{k}}Q)a_{3k}.
\end{eqnarray*}%
Next, integrating by parts the integrals containing $\overline{Q}_{j}$, we
obtain 
\begin{eqnarray*}
&&\frac{1}{2}\int\limits_{0}^{1}\int\limits_{0}^{2\pi }(t-\rho +2z\rho )\big(%
\nabla _{\xi }Q(\xi ,x^{0})\cdot \partial _{t}\xi \big)d\varphi dz \\
&=&\frac{\rho }{4}\int\limits_{0}^{1}\int\limits_{0}^{2\pi }(t-\rho +2z\rho
)\sum_{k,m=1}^{3}(\partial _{\xi _{k},\xi _{m}}^{2}Q)\Big\{\Big[z(1-z)\cos
\theta +\frac{(t-\rho +2z\rho )}{4}\sin ^{2}\theta \Big] \\
&&\times \lbrack b_{0km}+b_{1km}\cos (2\varphi )+b_{2km}\sin (2\varphi )] \\
&&+z(1-z)\sin \theta a_{3k}[(a_{1m}\cos \varphi +a_{2m}\sin \varphi
)+a_{3m}\cos \theta ]\Big\}d\varphi dz.
\end{eqnarray*}%
Moreover, 
\begin{eqnarray*}
&&\lim_{t\rightarrow |x_{x}^{0}|^{+}}\frac{1}{2}\int\limits_{0}^{1}\int%
\limits_{0}^{2\pi }(t-\rho +2z\rho )\big(\nabla _{\xi }Q(\xi ,x^{0})\cdot
\partial _{t}\xi \big)d\varphi dz\qquad \qquad \qquad \\
&=&\pi \rho ^{2}\int\limits_{0}^{1}z^{2}(1-z)(\Delta _{\xi
}Q)(x^{0}+z(x-x^{0}),x^{0})dz.
\end{eqnarray*}%
Hence, the first term in the right hand side of (\ref{3.20}) is a bounded
and continuous function in $\overline{G(x^{0},T)}$ for $n=2$ and it vanishes
for $n>2$. The second term in the right hand side of (\ref{3.20}) is also a
bounded and continuous function in $\overline{G(x^{0},T)}$ for $n=2,3$ and
vanishes for $n>3$. Since it was proven above that $\partial
_{t}^{2}w_{1}\in C\left( \overline{G(x^{0},T)}\right) ,$ then the third
intergal in the right hand side of (\ref{3.20}) is also a bounded and
continuous function in $\overline{G(x^{0},T)}$ for $n=2$. Using estimate (%
\ref{300}) and the obvious inequalities 
\begin{equation*}
|\partial _{\xi _{k},\xi _{m}}Q|\leq Cq_{0}q_{2}\leq Cq_{0}q_{4},\quad
k,m=1,2,3,
\end{equation*}%
we obtain 
\begin{equation*}
|\partial _{t}^{2}w_{2}(x,x^{0},t)|\leq Cq_{0}q_{4}.
\end{equation*}%
Next, using (\ref{3.20}), we obtain a similar estimate for the second
derivative of $w_{3}(x,x^{0},t)$: 
\begin{equation*}
|\partial _{t}^{2}w_{3}(x,x^{0},t)|\leq Cq_{0}^{2}q_{4}.
\end{equation*}%
Continuing theaw estimates, we obtain 
\begin{eqnarray*}
|\partial _{t}^{2}w_{4}(x,x^{0},t)| &\leq &\frac{Tq_{0}}{4\pi \rho }%
\int\limits_{\rho }^{t}d\tau \dint\limits_{E(x,x^{0},\tau
)}Cq_{0}^{2}q_{4}d\varphi dr=\frac{CTq_{0}^{3}q_{4}}{2}(t-\rho ), \\
|\partial _{t}^{2}w_{5}(x,x^{0},t)| &\leq &\frac{Tq_{0}}{4\pi \rho }%
\int\limits_{\rho }^{t}d\tau \dint\limits_{E(x,x^{0},\tau )}\frac{%
CTq_{0}^{3}q_{4}}{2}\Big(t-\frac{\tau +\rho }{2}\Big)d\varphi dr\leq \frac{%
CT^{2}q_{0}^{4}q_{4}}{2}\frac{(t-\rho )^{2}}{2!}.
\end{eqnarray*}%
Using the mathematical induction method, we get the following estimate 
\begin{equation*}
|\partial _{t}^{2}w_{n}(x,x^{0},t)|\leq \frac{CT^{n-3}q_{0}^{n-1}q_{4}}{2}%
\frac{(t-\rho )^{n-3}}{(n-3)!},\quad n\geq 4.
\end{equation*}%
Hence, all functions $\partial _{t}^{2}w_{n}\in C\left( \overline{G(x^{0},T)}%
\right) ,$ the series 
\begin{equation*}
\sum\limits_{n=1}^{\infty }\partial _{t}^{2}w_{n}(x,x^{0},t)
\end{equation*}%
converges uniformly in $\overline{G(x^{0},T)}$ and its sum $\partial _{t}^{2}%
\tilde{w}\in C\left( \overline{G(x^{0},T)}\right) $. $\square \hfill $

\end{document}